\documentclass[aps,prl,twocolumn,superscriptaddress,showpacs]{revtex4}

\usepackage{bm,graphics}
\begin{document}


\title{Does the $\boldsymbol{\Sigma(1580){3\over 2}^-}$ resonance exist?}


\author{J. Olmsted}
\altaffiliation{Present address: Midwest Proton Radiotherapy Institute,
2425 N. Milo Sampson Lane, Bloomington, IN 47408.}
\affiliation{Kent State University, Kent, OH 44242-0001}

\author{S.~Prakhov}
\affiliation{University of California Los Angeles, Los Angeles, CA 90095-1547}

\author{D.~M.~Manley}
\affiliation{Kent State University, Kent, OH 44242-0001}


\author{C.~E.~Allgower}
\altaffiliation{Present address: Midwest Proton Radiotherapy Institute,
2425 N. Milo Sampson Lane, Bloomington, IN 47408.}
\affiliation{Argonne National Laboratory, Argonne, IL 60439-4815}

\author{V.~S.~Bekrenev}
\affiliation{Petersburg Nuclear Physics Institute, Gatchina, Russia 188350}

\author{W.~J.~Briscoe}
\affiliation{The George Washington University, Washington, D.C. 20052-0001}

\author{M.~Clajus}
\affiliation{University of California Los Angeles, Los Angeles, CA 90095-1547}

\author{J.~R.~Comfort}
\author{K.~Craig}
\affiliation{Arizona State University, Tempe, AZ 85287-1504}

\author{D.~Grosnick}
\affiliation{Valparaiso University, Valparaiso, IN 46383-6493}

\author{D.~Isenhower}
\affiliation{Abilene Christian University, Abilene, TX 79699-7963}

\author{N.~Knecht}
\affiliation{University of Regina, Saskatchewan, Canada S4S~OA2}

\author{D.~D.~Koetke}
\affiliation{Valparaiso University, Valparaiso, IN 46383-6493}

\author{N.~G.~Kozlenko}
\author{S.~Kruglov}
\author{A.~A.~Kulbardis}
\affiliation{Petersburg Nuclear Physics Institute, Gatchina, Russia 188350}

\author{G.~Lolos}
\affiliation{University of Regina, Saskatchewan, Canada S4S~OA2}

\author{I.~V.~Lopatin}
\affiliation{Petersburg Nuclear Physics Institute, Gatchina, Russia 188350}

\author{R.~Manweiler}
\affiliation{Valparaiso University, Valparaiso, IN 46383-6493}

\author{A.~Maru\v{s}i\'{c}}
\altaffiliation{Present address: Collider-Accelerator Dept.,
Brookhaven National Laboratory, Upton, NY 11973.}
\author{S.~McDonald}
\altaffiliation{Present address: TRIUMF, 
4004 Wesbrook Mall, Vancouver, B.C., Canada V6T 2A3.}
\author{B.~M.~K.~Nefkens}
\affiliation{University of California Los Angeles, Los Angeles, CA 90095-1547}

\author{Z.~Papandreou}
\affiliation{University of Regina, Saskatchewan, Canada S4S~OA2}

\author{D.~C.~Peaslee}
\affiliation{University of Maryland, College Park, MD 20742-4111}

\author{N.~Phaisangittisakul}
\author{J.~W.~Price}
\affiliation{University of California Los Angeles, Los Angeles, CA 90095-1547}

\author{A.~F.~Ramirez}
\affiliation{Arizona State University, Tempe, AZ 85287-1504}

\author{M.~Sadler}
\affiliation{Abilene Christian University, Abilene, TX 79699-7963}

\author{A.~Shafi}
\affiliation{The George Washington University, Washington, D.C. 20052-0001}

\author{H.~Spinka}
\affiliation{Argonne National Laboratory, Argonne, IL 60439-4815}

\author{T.~D.~S.~Stanislaus}
\affiliation{Valparaiso University, Valparaiso, IN 46383-6493}

\author{A.~Starostin}
\affiliation{University of California Los Angeles, Los Angeles, CA 90095-1547}

\author{H.~M.~Staudenmaier}
\affiliation{Universit\"{a}t Karlsruhe, Karlsruhe, Germany 76128}

\author{I.~I.~Strakovsky}
\affiliation{The George Washington University, Washington, D.C. 20052-0001}

\author{I.~Supek}
\affiliation{Rudjer Bo\v{s}kovi\'{c} Institute, Zagreb, Croatia 10002}

\author{W.~B.~Tippens}
\altaffiliation{Present address: Nuclear Physics Div., 
Dept.\ of Energy, 19901 Germantown Road, Germantown, MD  20874-1290.}
\affiliation{University of California Los Angeles, Los Angeles, CA 90095-1547}

\collaboration{Crystal Ball Collaboration}
\noaffiliation


\date{\today}

\begin{abstract}
Precise new data for the reaction
$K^- p \rightarrow \pi^0 \Lambda$ are presented in the c.m.\ energy
range 1565 to 1600~MeV.  
Our analysis of these data
sheds new light on claims for the $\Sigma(1580)\frac{3}{2}^-$ resonance,
which (if it exists with the specified quantum numbers) must be an exotic baryon because of its
very low mass.  Our results 
show no evidence for this state.
\end{abstract}


\maketitle

The $\Sigma(1580)\frac{3}{2}^-$ resonance was first reported in 1973
as a narrow $I=1$
structure in a preliminary analysis of
high statistics measurements of $K^- p$ and $K^- d$
total cross sections \cite{Li73}.  It was subsequently observed
in $K^- p \rightarrow \pi^0 \Lambda$ by Litchfield who identified it
as a $\frac{3}{2}^-$ state with mass
$1582 \pm 4$~MeV, width $11 \pm 4$~MeV, and amplitude at resonance
$0.10 \pm 0.02$ \cite{Litchfield74}.  
For his analysis, Litchfield used the only experimental data then
available in the c.m.\ energy range 1560 to 1600~MeV, which were the
bubble-chamber measurements of Armenteros {\it et al.}\ \cite{Armenteros70}.
Some details of his analysis will be presented below.
The final analysis of the
total $K^-$-nucleon total cross sections yielded mass
$1583 \pm 4$~MeV, width 15~MeV, and $(J+1/2)~x = 0.06$, where $x$
is the ${\overline{K}}N$ branching fraction \cite{Carroll76}.  

The $\Sigma(1580)\frac{3}{2}^-$ resonance, if confirmed,
would be of strong theoretical
interest because it cannot be accommodated as an ordinary
$q^3$ state due to its very low mass. 
More specifically, this state would necessarily be the lowest $\Sigma^*$  with
$J^P = \frac{3}{2}^-$.  One expects such a state to be the octet partner
of the $N(1520)\frac{3}{2}^-$.  However, that partner is well established
as the $\Sigma(1670)\frac{3}{2}^-$, in good agreement with quark-model
predictions \cite{Capstick86}.  The next $\Sigma^*$ with the right quantum
numbers is expected to lie about 100~MeV {\it higher} than the
$\Sigma(1670)\frac{3}{2}^-$.  

Two bubble-chamber experiments \cite{Engler76,Cameron78} that studied the analog reaction 
$K^0_L p \to \pi^+ \Lambda$ were performed subsequent to Litchfield's
analysis.  The results of these two experiments
did not {\it require} the $\Sigma(1580)\frac{3}{2}^-$ but neither was it
ruled out.
In this Letter, we present precise new experimental results of the
$K^- p \rightarrow \pi^0 \Lambda$ reaction in the c.m.\ energy
range 1565 to 1600~MeV.  


Our measurements of the $K^-p\to \pi^0\Lambda$ reaction
 were conducted at Brookhaven National Laboratory (BNL)
 with the Crystal Ball (CB) multiphoton spectrometer, which was installed in
 the C6 beam line of the Alternating Gradient Synchrotron \cite{Olmsted01}.
 The experiment was performed with a momentum-analyzed $K^-$ beam
 incident on a 10-cm-long liquid hydrogen target located in
 the center of the CB.  While
measurements were performed at eight different beam momenta, 
this Letter reports only
on the lowest three momenta: 514, 560, and 581~MeV/$c$.
We plan a more extensive report on the full data set at a later date.

The Crystal Ball consists of 672 optically isolated NaI(Tl) crystals,
 shaped like truncated triangular pyramids and
arranged in two hemispheres that cover 93\% of $4\pi$ steradians.
The event trigger was a coincidence of the
beam trigger and a CB signal that
required the total energy deposited in the crystals
 to exceed some threshold. 
Our trigger for neutral events required, 
in addition, that there was no corresponding signal 
 from the system of scintillation counters surrounding
 the target.  
 We achieved a typical energy resolution for electromagnetic showers in the CB
 of $\Delta E/E = 0.020/E[{\rm GeV}]^{0.36}$.
 Shower directions were measured with a resolution in $\theta$,
 the polar angle with respect to the beam axis, of
$\sigma_\theta = 2^\circ \textrm{--} 3^\circ$ for photon energies
 in the range 50 to 500~MeV, assuming that the photons were
 produced in the hydrogen target.
 The resolution in $\phi$ was $2^\circ/\sin\theta$.
 Our resolution for the momentum determination of
 individual beam particles was about 0.6\%, whereas
the uncertainty of 
the mean momentum of the beam on target
 was about 0.3\%.
 The momentum divergence of the kaon beam on target
 was about 2\%. 
To illustrate the CB mass resolution, we show in
 Figs.~1a-c the invariant mass spectra
 for the experimental events that
 were selected by kinematic fit to three hypotheses:
 $K^-p \to \pi^0\pi^0\pi^0\Lambda$,
 $K^-p \to \gamma\gamma\Lambda$,
 and $K^-p \to \pi^0\gamma\Lambda$.
 The $\Lambda$ hyperon was identified by its decay into $\pi^0 n$.
 The spectra shown were obtained for a beam momentum of 750~MeV/$c$,
 which is above the threshold for $K^-p \to \eta\Lambda$.
 The invariant mass of the $3\pi^0$ peak from $\eta$ decays has
 a root-mean-square width
 $\sigma_m(3\pi^0) \approx 5$~MeV in comparison with
 $\sigma_m(2\gamma) \approx 6$~MeV for $\eta\to\gamma\gamma$.
 Since the constraint on the $\Lambda$ hyperon mass was used in the
 kinematic fit for the secondary vertex determination, we cannot
 illustrate the $\Lambda$ mass resolution. Instead, we 
 show in Fig.~1c the $\gamma\Lambda$ peak from $\Sigma^0$ decays,
 which has $\sigma_m(\gamma\Lambda) \approx 6$~MeV.
 More details on the CB apparatus, trigger conditions, and the absolute
 cross-section determination
 can be found in Refs.~\cite{etalam,eta_slope,nakorn}.

 To select candidates for the reaction
\begin{equation}
\label{eqn:lampi0}
K^-p \to \pi^0\Lambda \to \pi^0\pi^0 n \to 4\gamma n~,
\end{equation}
 we used the neutral 4- and 5-cluster events. 
A ``cluster'' is a group of neighboring crystals in which energy is
deposited from a single-photon electromagnetic shower.
The neutral clusters have a 20-MeV threshold in software
and were analyzed as single photons.
 We assumed that all photons
 produced electromagnetic showers in the CB and that the neutron
 was either the missing particle (for 4-cluster events)
 or that it was detected by the CB.

 The hypothesis of reaction~(\ref{eqn:lampi0})
 was tested by a kinematic fit for
 all possible pairing combinations of four photon clusters
 to yield two $\pi^0$'s, with one of them being the $\pi^0$ from
 the $\Lambda$ decay.
 The measured parameters in the kinematic fit included
 five for the beam particle (momentum, $\theta_X$ and $\theta_Y$ angles,
 and $X$ and $Y$ coordinates on the target) and three
 for each photon cluster (energy, $\theta$ and $\phi$ angles).
 When the neutron was the missing particle, its
 energy and $\theta$ and $\phi$ angles were free parameters of
 the fit. If the neutron was detected by the CB, its
 $\theta$ and $\phi$ angles were used as the measured parameters.
 The $Z$ coordinate of the primary vertex and
 the decay length of the $\Lambda$ were also free parameters
 of the kinematic fit. Since we have three constraints 
on the masses of the two $\pi^0$'s
 and the $\Lambda$, in addition to the four main
 constraints of the kinematic fit (energy and 3-momentum conservation),
the total number of constraints
 in the hypothesis of reaction~(\ref{eqn:lampi0}) is seven.
 The effective number of constraints is this total less the number
 of free parameters of the fit; it results in  2-C and
 4-C fits for 4- and 5-cluster events, respectively.
 In the case of 4-cluster events, we have three possible pairing combinations
 to yield the two $\pi^0$'s; this becomes six combinations when
we further allow the choice of each $\pi^0$ to be from the $\Lambda$ decay.
 For 5-cluster events, the number of pairing combinations is
 five times larger, as the neutron is also involved
 in the cluster permutations.

 Those events in which at least one pairing combination
  satisfied our hypothesis at the 5\% C.L. ({\it i.e.}, with probability
 $>$~5\%) were accepted for further analysis.
 The pairing combination
 with the largest confidence level was used to reconstruct
 the kinematics of the reaction.
 Additional selection criteria (on the primary vertex,
 the decay length of the $\Lambda$, and confidence level of other
 possible hypotheses) were also applied for better suppression
 of background processes.
 The largest background turned out to be from
 processes that were not kaon beam interactions in the liquid
 hydrogen target. The major part of these background events were
 $K^-$ decays in the beam. Such events were analyzed
 with ``empty-target'' samples and then were
 subtracted from the experimental distributions. 
These events did not survive the kinematic fit C.L. plus other
selection criteria.
The fraction of
 this background subtracted from the selected
 events of reaction~(\ref{eqn:lampi0}) comprises 6--9\%.

 Another significant background could arise from the misidentification
 of ~$K^-p \to K^0_S n \to \pi^0\pi^0 n \to 4\gamma n$~ events
 as our $K^-p\to \pi^0\Lambda$ reaction. To suppress this
 background, we fitted our events to both the $K^-p\to \pi^0\Lambda$
 and $K^-p \to K^0_S n$ hypotheses and applied the following selection
 criterion on the largest probabilities of the kinematic fits:
 ~$P(K^-p\to \pi^0\Lambda) > 2\times P(K^-p\to K^0_S n)$.
 This enabled us to decrease the $K^0_S n$ background
 to a level less than 4\%.
 Background from
 ~$K^-p \to \pi^0\Sigma^0 \to \pi^0\gamma\Lambda \to 5\gamma n$ events
 was found to be less than 2\%.
For one momentum, we simulated large statistics for the $\pi^0 \Sigma^0$
and $K^0_S n$ backgrounds according to their differential cross sections.
The subtraction of these backgrounds with their small weights did not
change the shape of the $K^- p \to \pi^0 \Lambda$ differential cross
sections, it was just made smaller by the percent of the events subtracted;
consequently, 
for the other momenta, the $K^- p \to \pi^0 \Lambda$ differential cross
sections were just corrected by the fraction of $\pi^0 \Sigma^0$
and $K^0_S n$ backgrounds, without the direct subtraction as was done
for the empty-target background.

 The experimental number of the selected events
 was 3539, 6741, and 10046 for beam momenta of
 514, 560, and 581~MeV/$c$, respectively. About 20\% of these events
 were reconstructed from the 5-cluster sample ({\it i.e.},
 with the neutron detected in the CB).
 The combinatorial background in our $K^-p\to \pi^0\Lambda$ events
 was estimated to be not larger than 1--2\%.
Reconstructed $\pi^0\Lambda$ events 
comprised about 0.2\% of all  
neutral-trigger events collected in the given momentum range.

 Acceptance corrections were based on
 a Monte Carlo simulation of
 the $K^-p \to \pi^0\Lambda \to \pi^0\pi^0 n$
 reaction for each momentum,
 which was performed according to phase space by using
 the experimental beam-trigger events as input for kaon beam
 distributions. Specifically, the phase-space Monte Carlo
simulation produced the $\pi \Lambda$ state 
with an isotropic angular distribution
with respect to the beam direction; further,
the decay of the $\Lambda$ in its rest frame was also assumed to be
isotropic.
The CB neutral trigger and resolution conditions
 for data sets of each beam momentum were properly
 reproduced by the Monte Carlo simulations. 
The simulated and experimental distributions for the $\Lambda$ decay length
were in good agreement, and comparable to those shown in 
Fig.~12 of Ref.~\cite{etalam}, which shows the $\Lambda$ decay-distance
distributions for $K^- p \to \eta \Lambda$.  Our Monte Carlo simulations
used the PDG decay length $c\tau=7.89$~cm.  From our simulation
of $K^- p \to \pi^0 \Lambda$ events, we determined that the
primary vertex resolution was $\sigma_z = 1.9$~cm,
the $\Lambda$ decay length resolution was $\sigma_{\rm decay} = 3.0$~cm,
and the c.m.\ $\theta$ resolution for outgoing neutral pions was
$\sigma_\theta \approx 0.06$~rad (or $\approx 3.4^\circ$).
Our mean detection efficiency
 for the $K^-p \to \pi^0\Lambda \to \pi^0\pi^0 n$ events
 was about 28\%.
The typical CB acceptance for the $\cos\theta$ distribution
is shown in Fig.~1d.
 For the calculation of the cross sections, we used
 the PDG branching ratio for the $\Lambda \to \pi^0 n$ decay
 of 0.358$\pm$0.005 \cite{PDG}. The effective proton density
 in the target times the effective target length
was $(4.05\pm0.08)\times 10^{-4}$~mb$^{-1}$.
 An additional systematic uncertainty in the
 obtained cross sections, which
comes mostly from the calculation of the kaon beam monitor 
number and from the evaluation
 of the fraction of useful events lost due to pile-up in the CB,
was estimated to be less
 than 7\%.  This contribution was not included in the
 quoted final uncertainty.  The pile-up events (clusters) were easily
removed using the measured TDC information.

The polarization of the $\Lambda$ was measured via its decay
asymmetry:
\begin{equation}
P(\cos{\theta}) = 3 \left ( \sum_{i=1}^{N(\theta)} \cos{\xi_i} \right ) /
(\alpha_\Lambda N(\theta))~.
\end{equation}
Here $\theta$ refers to the direction of the outgoing $\pi^0$ (not
from $\Lambda$ decay) with respect to the incident $K^-$ meson in the
c.m.\ system.  $N(\theta)$ is the total number of $K^- p \to
\pi^0 \Lambda$ events in the $\cos{\theta}$ bin, and $\alpha_\Lambda =
+0.65$ is the $\Lambda$ decay asymmetry parameter.
The angle $\xi$ was defined for the $i$-th event
in the $\cos{\theta}$ bin by 
$\cos{\xi} = \hat{{\boldsymbol{N}}}\cdot \hat{{\boldsymbol{n}}}$, 
where $\hat{{\boldsymbol{n}}}$ is a unit vector in the direction
of the decay neutron (in the $\Lambda$ rest frame) and 
$\hat{{\boldsymbol{N}}}$ is a unit vector normal to the production plane:
$\hat{\boldsymbol{N}} = ({\hat {\boldsymbol{K}}}^- \times {\hat{\boldsymbol{\pi}}}^0)/
 |{\hat {\boldsymbol{K}}}^- \times {\hat{\boldsymbol{\pi}}}^0|.$
Finally,
the vectors ${\hat{\boldsymbol{K}}}^-$ and ${\hat{\boldsymbol{\pi}}}^0$
are unit vectors in the direction of the incident $K^-$ and the
outgoing $\pi^0$ meson, respectively.


Our Monte Carlo simulations for $K^- p \to \pi^0 \Lambda$ were
generated with no polarization for the produced $\Lambda$; however,
simulations with a nonzero 
$\Sigma^0$ polarization were performed for the reaction
$K^- p \to \pi^0 \Sigma^0$ to investigate possible acceptance effects in the
$\Sigma^0$ polarization.  These studies suggest a possible acceptance effect
of about 10-15\% of the polarization value. 
It is notable that our polarization measurements are in
good agreement \cite{CB-01-010} 
with the bubble-chamber results of Armenteros {\it et al.} 
 \cite{Armenteros70} where both
experiments have enough statistics ({\it e.g.}, at 750 MeV/$c$) to make a 
sensible comparison.  One expects acceptance effects to be very
small in bubble-chamber experiments. For the results discussed in this Letter,
statistical uncertainties were the main contributions to the
total uncertainties in our  $\Lambda$ polarization measurements.

For the comparisons required in this work, it was necessary to rebin our data into
smaller momentum intervals.  
This was made possible because
we measured the deviation from the nominal beam momentum for each event. 
The size of these bins was limited by the precision of the relative
$K^-$ momentum determination, which was $\pm 2$~MeV/$c$ \cite{etalam}.
Data at the beam momentum 514~MeV/$c$ were rebinned into two
intervals with mean momenta of 506 and 523~MeV/$c$.  Data at 560~MeV/$c$
were rebinned into three intervals (544, 560, and 576~MeV/$c$), and
data at 581~MeV/$c$ were rebinned into two intervals (571 and 591~MeV/$c$). 
Figure~\ref{data_560_591} shows our representative 
data for $d\sigma/d\Omega$
and the $\Lambda$ polarization in $K^- p \rightarrow \pi^0 \Lambda$
at 560 and 591~MeV/$c$. Our measurements 
are in good agreement with older measurements \cite{Armenteros70},
although our results are of much higher statistical precision.
In particular, our results for the $\Lambda$ polarization
are a dramatic improvement over older measurements \cite{Armenteros70}.
Additional illustrations of the good agreement of our results with
older measurements can be found in a CB internal report \cite{CB-01-010}.

\begin{figure}[h]
\scalebox{0.45}{\includegraphics*{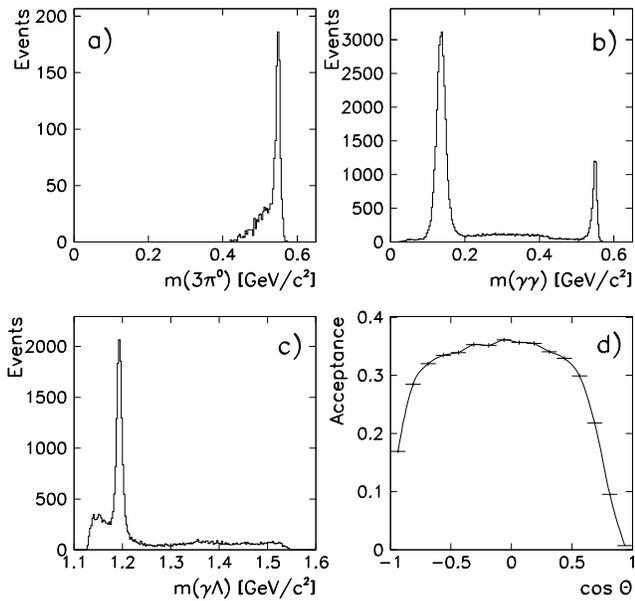}}
\caption{\small The invariant mass distributions for the experimental events
 at $p_{K^-} = 750$~MeV/$c$ that
 were selected by kinematic fit to the three hypotheses:
 ~(a) $K^-p \to \pi^0\pi^0\pi^0\Lambda \to 4\pi^0 n \to 8\gamma n$,
  $\sigma_m(\eta\to 3\pi^0) \approx 5$~MeV;
 ~(b) $K^-p \to \gamma\gamma\Lambda \to \gamma\gamma\pi^0 n \to 4\gamma n$,
  $\sigma_m(\eta\to 2\gamma) \approx 6$~MeV,
  $\sigma_m(\pi^0\to 2\gamma) \approx 13$~MeV;
 and ~(c) $K^-p\to \gamma\pi^0\Lambda\to \gamma\pi^0\pi^0 n\to 5\gamma n$,
 $\sigma_m(\Sigma^0\to \gamma\Lambda) \approx 6$~MeV.
 (d) Acceptance of $K^- p \to \pi^0\Lambda$ for
 the $\cos \theta$ distribution at $p_{K^-} = 560$~MeV/$c$,
 where $\theta$ refers to the direction of the outgoing $\pi^0$
 with respect to the incident $K^-$ meson in c.m. system.}
\label{distributions} 
\end{figure}

\begin{figure}[h]
\scalebox{0.50}{\includegraphics*{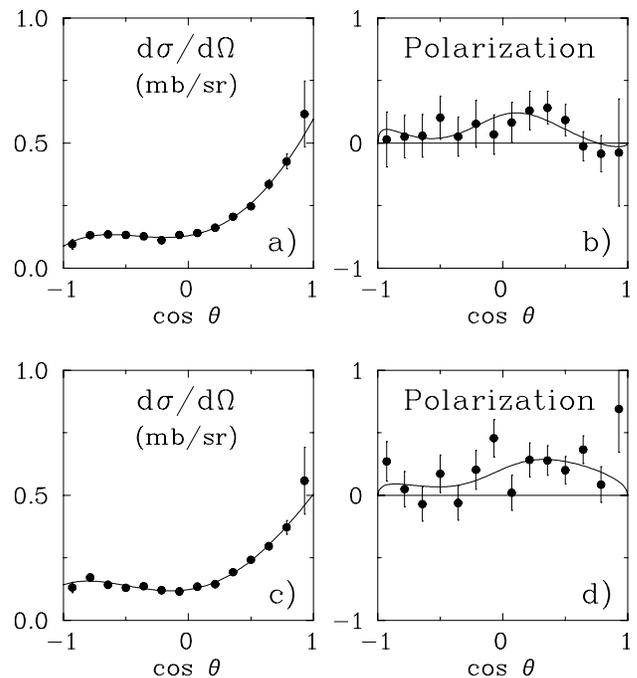}}
\caption{\small Top: Differential cross section (a) and the induced $\Lambda$ polarization (b) for $K^- p \rightarrow \pi^0 \Lambda$ at 560~MeV/$c$.  
Bottom: Differential cross section (c) and the induced $\Lambda$ polarization (d) at 591~MeV/$c$. Only statistical uncertainties are shown. 
The curves show the results of fitting our data by using the expansions in
Eqs.~(3) and (4) with terms up to $L=4$, as described in
the text.}
\label{data_560_591} 
\end{figure}


The differential and transversity cross sections
($\Lambda$ polarization
times differential cross section) were expanded in terms of
Legendre polynomials and associated Legendre functions, in the usual way: 
\begin{eqnarray}
\frac{d\sigma}{d\Omega} = \lambdabar^2 A_0 \sum_{L=0}^{N} (A_L / A_0) P_L (\cos \theta )~, \\
P \frac{d\sigma}{d\Omega} = \lambdabar^2 A_0 \sum_{L=1}^{N} (B_L / A_0) P_L^1 (\cos \theta )~. 
\end{eqnarray}
Coefficients with $L>4$ were consistent with zero.

\begin{figure}
\scalebox{0.45}{\includegraphics*{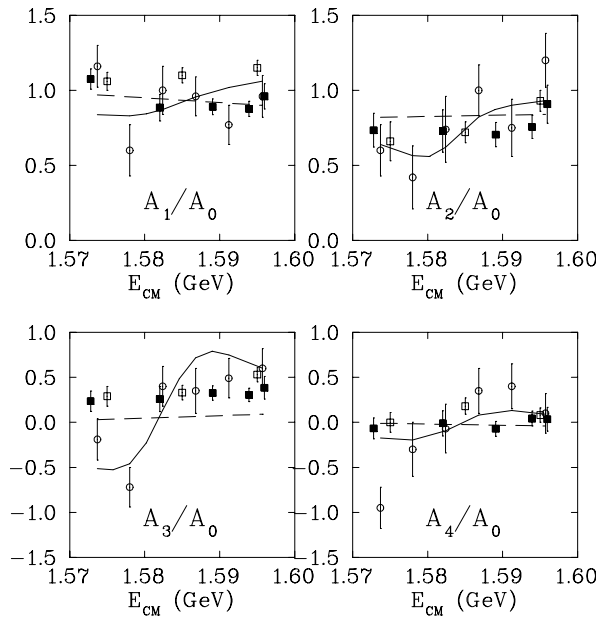}}
\caption{\small $A_L/A_0$ coefficients as a function of c.m.\ energy
(see text).}
\label{A1_A0} 
\end{figure}

\begin{figure}
\scalebox{0.45}{\includegraphics*{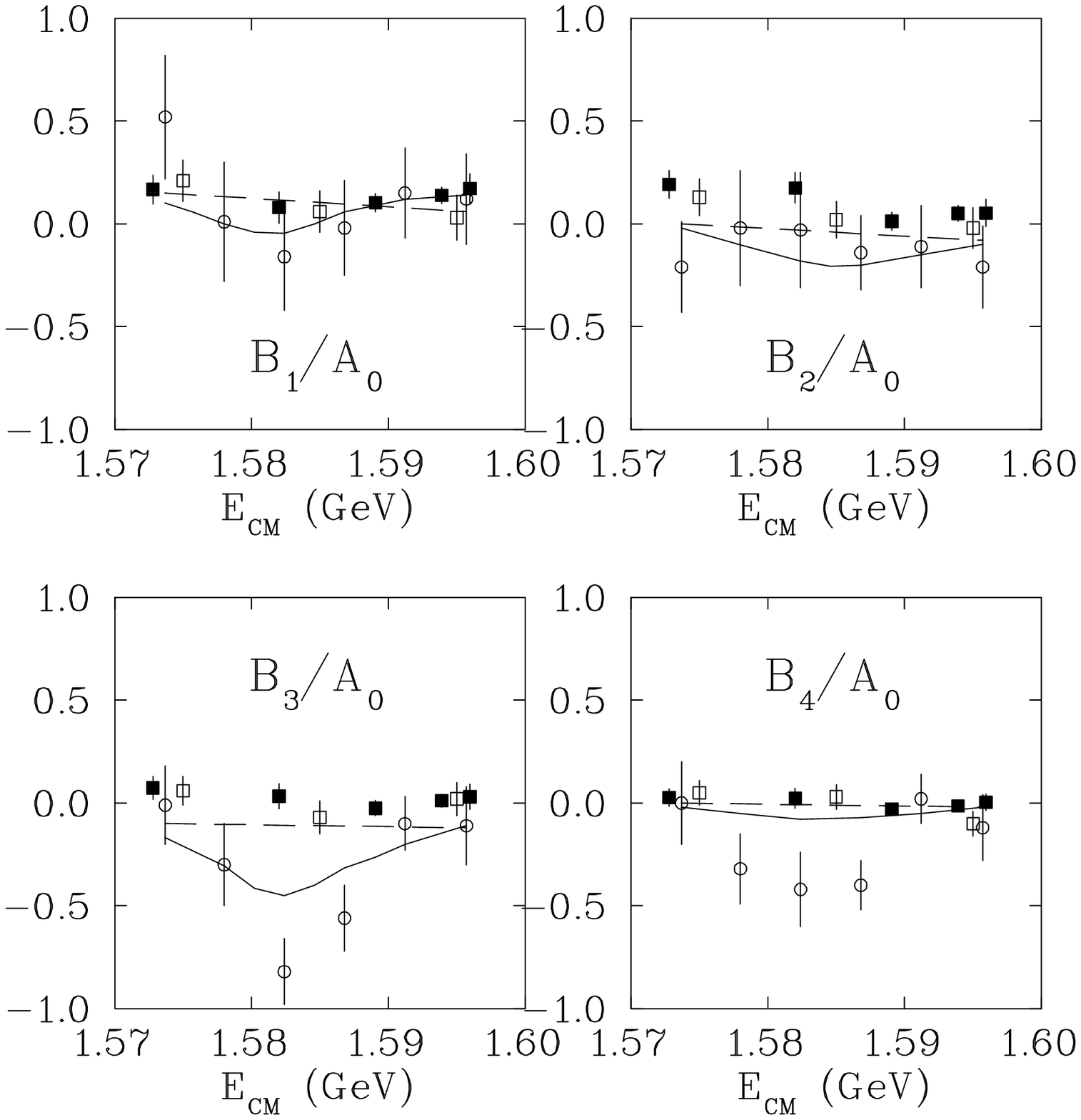}}
\caption{\small $B_L/A_0$ coefficients as a function of c.m.\ energy (see text).}
\label{B1_A0} 
\end{figure}

Figures~\ref{A1_A0} and \ref{B1_A0}, respectively, show 
experimentally determined values (open circles) of the
$A_L/A_0$ and $B_L/A_0$ coefficients from the analysis of
Litchfield \cite{Litchfield74}.  Our new results are shown
by solid squares.   
Open squares denote the $K^0_L p \to \pi^+ \Lambda$ 
measurements by Cameron {\it et al.} \cite{Cameron78}.
Older $K^0_L p \to \pi^+ \Lambda$
results by Engler {\it et al.}
\cite{Engler76} are not shown since they have comparable
precision to those from the Litchfield analysis and do little
to resolve the issue.
Table~\ref{Comparison}
gives our results for the fitting coefficients shown in Figs.\ \ref{A1_A0} and \ref{B1_A0}.
The dashed curves in the figures show the
results of Litchfield's first fit to the 
measured coefficients determined from the rebinned data of
Armenteros {\it et al.} \cite{Armenteros70}.
This fit was made with constant S11, P11, and P13 amplitudes
together with the tails of the D13 $\Sigma(1660)$ and D15 $\Sigma(1765)$
resonances.  The solid curves show Litchfield's second fit,
which added a resonant D13 amplitude.  
The parameters of the added
D13 resonance were found to be $M = 1582 \pm 4$~MeV, 
$\Gamma = 11 \pm 4$~MeV, and $\sqrt{xx^\prime} = 0.10 \pm 0.02$.
It is evident from these figures that our results disagree 
strikingly with
the solid curves that reflect inclusion of the $\Sigma(1580)\frac{3}{2}^-$
resonance.  More generally, our results show no rapid variation
to suggest the presence of 
{\it any} narrow resonance in the c.m.\ energy range 1570 to 1600~MeV.

\begin{table*}
\caption{Fitting coefficients for the low-momentum $d\sigma/d\Omega$ and $P \cdot d\sigma/d\Omega$ data.
The incident momentum $P_{\textrm{lab}}$ and c.m. energy $E_{\textrm{\tiny CM}}$ are in MeV/$c$ and MeV, respectively.}
\label{Comparison}
\begin{ruledtabular}
\begin{tabular}{rrrrrrrrrrr}
\multicolumn{1}{c}{$P_{\textrm{lab}}$} & \multicolumn{1}{c}{$E_{\textrm{\tiny CM}}$} & \multicolumn{1}{c}{$A_0$}  & \multicolumn{1}{c}{$A_1/A_0$} & \multicolumn{1}{c}{$A_2/A_0$} & \multicolumn{1}{c}{$A_3/A_0$} & \multicolumn{1}{c}{$A_4/A_0$} & \multicolumn{1}{c}{$B_1/A_0$} & \multicolumn{1}{c}{$B_2/A_0$} & \multicolumn{1}{c}{$B_3/A_0$} &\multicolumn{1}{c}{$B_4/A_0$} \vspace{1 ex}\\ \colrule
506 & 1565 &  0.0535(22) & 1.157(66) & 0.77(11) & 0.29(11) & $-$0.00(16) & 0.132(68) & 0.078(66) & 0.006(56) & 0.017(39)\\
523 & 1573 &  0.0512(21) & 1.075(68) & 0.73(11) & 0.24(11) & $-$0.07(12) & 0.168(70) & 0.192(68) & 0.074(56) & 0.027(42)\\
544 & 1582 &  0.0584(30) & 0.885(89) & 0.73(14) & 0.26(14) & $-$0.01(14) & 0.080(76) & 0.175(74) & 0.033(60) & 0.022(48) \\
560 & 1589 &  0.0588(18) & 0.890(53) & 0.70(8)  & 0.33(8)  & $-$0.07(8)  & 0.103(45) & 0.012(43) & $-$0.025(37) & $-$0.029(28) \\ 
571 & 1594 &  0.0626(18) & 0.876(50) & 0.76(8)  & 0.30(7)  &    0.04(8)  & 0.139(39) & 0.051(38) & 0.011(32) & $-$0.014(25)\\
576 & 1596 &  0.0795(38) & 0.960(84) & 0.91(13) & 0.38(13) &    0.03(13) & 0.171(72) & 0.053(67) & 0.030(61) & 0.004(39) \\
591 & 1603 &  0.0618(18) & 0.684(54) & 0.70(8)  & 0.22(8)  & $-$0.10(8)  & 0.196(41) & 0.101(39) & 0.045(33) & $-$0.005(26)\\
\end{tabular}
\end{ruledtabular}
\end{table*}


We have presented precise new results for the reaction $K^- p \to \pi^0 \Lambda$ spanning the c.m.\ energy range 1565 to 1600~MeV. 
Our results are compared with older data and with the
partial-wave analysis that introduced the
$\Sigma(1580)\frac{3}{2}^-$ resonance. One may use flavor-symmetry arguments or quark-model calculations
to rule out such a state as an ordinary $q^3$ baryon. 
Our measurements, especially when taken together with the earlier
results of Cameron  {\it et al.}\ \cite{Cameron78},
find no evidence for the $\Sigma(1580)\frac{3}{2}^-$ state.


\begin{acknowledgments}
This work was
supported in part by DOE, NSF, NSERC of Canada, and the Russian
Ministry of Science and Technology.  We thank SLAC for the loan
of the Crystal Ball.  The assistance of BNL is also
gratefully acknowledged.
\end{acknowledgments}


\end{document}